\documentclass[A4,12pt]{article}

\usepackage{amsmath}

\newcommand{\hs}{\hspace}
\newcommand{\ts}{\hspace*}
\newcommand{\vs}{\vspace}
\newcommand{\e}{\enskip}
\newcommand{\q}{\quad}

\newcommand{\dps}{\displaystyle}
\newcommand{\f}{\frac}

\newcommand{\Phat}{\hat{{\cal P}}}

\newcommand{\Qhat}{\hat{\Omega}}

\newcommand{\hxi}{\hat{\xi}}
\newcommand{\hpi}{\hat{\pi}}

\newcommand{\Zp}{\hat{Z}^{(+)}}
\newcommand{\Zm}{\hat{Z}^{(-)}}
\newcommand{\Zpm}{\hat{Z}^{(\pm)}}
\newcommand{\Zmp}{\hat{Z}^{(\mp)}}

\newcommand{\hope}{{\it hyper}-operator}
\newcommand{\bM}{\bar{M}}
\newcommand{\mcK}{\mathcal{K}}
\newcommand{\mcC}{\mathcal{C}}
\newcommand{\mcA}{\mathcal{A}}

\newcommand{\mcS}{\mathcal{S}}

\newcommand{\mcG}{\mathcal{G}}

\newcommand{\tmcC}{\mathcal{C}^{(\mbtn{1})}}
\newcommand{\tmcK}{\mathcal{K}^{(\mbtn{1})}}
\newcommand{\tmcS}{\mathcal{S}^{(\mbtn{1})}}
\newcommand{\tphi}{\phi}
\newcommand{\tpsi}{\psi}
\newcommand{\tH}{H^{(\mbtn{1})}}

\newcommand{\commu}[2]{[#1,#2]}
\newcommand{\commut}[2]{[#1,\e #2]}
\newcommand{\symmp}[2]{\ \{#1,\e #2\}}
 
\newcommand{\mbtn}[1]{\mbox{{\tiny #1}}}
\newcommand{\pscrp}{\mbox{{\scriptsize $\Phat$}}}

\newcommand{\hum}[1]{\hat{u}_{#1}^{(-)}}

\newcommand{\hpxm}[1]{\hat{p}^{x(-)}_{#1}}
\newcommand{\hpum}[1]{\hat{p}_u^{#1(-)}}

\makeatletter
 
\@addtoreset{equation}{section}
\makeatother

\begin{document}

\begin{center}
{\bf \Large Alternative Approach to Noncommutative Quantum Mechanics on a Curved Space
}\vs{12pt}\\
M. Nakamura\footnote{\label{*}E-mail:mnakamur@hm.tokoha-u.ac.jp}
\vs{12pt}\\
{\it Research Institute, Hamamatsu Campus, Tokoha University, Miyakoda-cho 1230, 
Kita-ku, Hamamatsu-shi, Shizuoka 431-2102, Japan}
\end{center}

\begin{abstract}
Starting with the first-order singular Lagrangian containing the redundant variables, the  noncommutative quantum mechanics on a curved space is investigated by the constraint star-product quantization formalism of the projection operator method. Imposing the additional constraints to eliminate the reduntant degrees of freedom,  the noncommutative quantum system with noncommutativity among the coordinates on the curved space is {\it exactly} constructed.  Then, it is shown that the resultant Hamiltonian contains the quantum corrections in the {\it exact} form.  We further discuss the additional  constraints to realize the noncommutativities both of coordinates and momenta on the curved space.
\end{abstract}


\section{Introduction}	
\ts{12pt}The problem of the quantization of a dynamical system constrained to a submanifold   embedded in the higher-dimensional Euclidean space has been extensively investigated as one of the quantum theories on a curved space until now\cite{A1}, and the noncommutative extensions of  quantum mechanics and quantum field theories have been also done widely\cite{A2}. Then, the noncommutative extensions of  the quantum theories on the curved spaces have been investigated with much interest\cite{A3,A4}. As the curved space, the submanifold $M^{N-1}$ specified by $G(x)=0$ ($G(x)\in {\it C}^{\infty}$) in an $N$-dimensional Euclidean space $R^N$ has been considered in many studies, where $x=(x^1,\cdots,x^i,\cdots,x^N)\in R^N$. \\
\ts{12pt}For the dynamical system constrained to $M^{N-1}$, there have been considered the two kinds of constraint conditons, one of which is the {\it static} constraint, $G(x)=0$, which does not conserve the canonically conjugate commutation relations(CCR) on the flat space, and the others, the {\it dynamical} one, $\dot{G}(x)=0$, which does the CCR\cite{A1}. Then, it has been shown\cite{A4} that the noncommutativities of the system on the curved space provide the so-called quantum corrections in the Hamiltonian and the commutator algebra of the system, which are missed in the usual approach to the quantization of constraint systems. Therefore, it is extremely interesting to construct the noncommutative quantum mechanics constrained to $M^{N-1}$ in the {\it exact} form.\\
\ts{12pt}For this purpose, in this paper, we shall at first present the Faddeev-Jackiw type\cite{A5} first order singular Lagrangian with the {\it dynamical} constraint, which is expressed with the coordinates $x$, $v$, $\lambda$ and $u$, where $v=(v_1,\cdots,v_i,\cdots,v_N) \in R^N$,  $\lambda \in R$ associated to the constraint and  $u=(u_1,\cdots,u_i,\cdots,u_N)\in R^N$, the redundant variables\cite{A6} associated to the noncommutativities among the resultant canonically conjugate set of the system.\\
\ts{12pt}Starting with the Lagrangian mentioned above, we shall construct the noncommutative quantum mechanics on the curved space through the constraint star-product quantization formalism of the projection operator method (POM)\cite{A7,A8}. Imposing the additional constraints to eliminate the redundant degrees of freedom, then, the noncommutative quantum system with the noncommutativity among the coordinates $x$ on the curved space will be constructed {\it exactly}  and it will be shown in the {\it exact} form that the Hamiltonian contains the quantum correction terms due to the noncommutativity associated to the constraint-opetator $G(x)$ and the successive projections of the Hamiltonian. We shall further investigate the other additional constraints associated to the noncommutativities both of coordinates and momenta. \\
\ts{12pt}This paper is organized as follows. In Sect.2, we propose the brief review of the constraint  star-product quantization formalism of the POM. In Sect.3, we set up the initial unconstraint quantum system. Imposing the two kinds of addtional constraints to eliminate the redundant degrees of freedom, in Sect.4, the resultant noncommutative quantum systems on the curved space are constructed in the {\it exact} form, and the quantum corrections in these resultant systems are investigated. In Sect.5, the some concluding remarks are given. 

\section{Star-product Quantization}

\ts{12pt}Following our previous works\cite{A2,A7,A8}, we here present the brief review of the constraint star-product quantization formalism of the POM together with the explanation of the notations appearing hereafter.\\ 
\ts{12pt}Let $\mcS=(\mcC,\mcA(\mcC),H(\mcC),\mcK)$ be the initial unconstraint quantum system, where $\mcC=\{(q^i,p_i);i=1,\cdots ,N\}$ is a set of canonically conjugate operators (CCS), $\mcA(\mcC)$, the commutator algebra of $\mcC$ with  
$$
\mcA(\mcC):\commu{q^i}{p_j}=i\hbar\delta^i_j,\hs{36pt}\commu{q^i}{q^j}=\commu{p_i}{p_j}=0,
\eqno{(2.1)}
$$ 
and $H(\mcC)$ is the Hamiltonian of the initial unconstraint system, $\mcK=\{T_{\alpha}(\mcC)|\alpha=1,\cdots ,2M<2N\}$, the set of the constraint-operators $T_{\alpha}(\mcC)$ corresponding to the second-class constraints $T_{\alpha}\approx  0$. Starting with $\mcS$, our task is to construct the constraint quantum system  
$\mcS^*=(\mcC^*,\mcA(\mcC^*),H^*(\mcC^*))$, where  $\mcC^*$ is the projected CCS satisfying
$$
T_{\alpha}(\mcC^*)=0\hs{48pt}(\alpha=1,\cdots,2M).
\eqno{(2.2)}
$$
\ts{12pt}The first step is to construct the associated canonically conjugate set (ACCS) from the constraint-operators $T_{\alpha}(\mcC)$ and the projection operator $\Phat$ to eliminate $T_{\alpha}\ (\alpha=1,\cdots,2M)$. \\
\ts{12pt}Let $\{(\xi^a,\pi_ a)|a=1,\cdots,M\}$ be the ACCS, and their symplectic forms be
$$ 
Z_{\alpha}=\left\{\begin{array}{l}\xi^a\hs{24pt}(\alpha=a)\vs{6pt}\\
\pi_a\hs{24pt}(\alpha=a+M) \hs{36pt}(\alpha=1,\cdots,2M\q;\q a=1,\cdots,M),
\end{array}\right.
\eqno{(2.3)}
$$
which obey the commutator algebra
$$
\begin{array}{l}
\commut{\xi^a}{\pi_b}=i\hbar\delta^a_b,\hs{12pt}\commut{\xi^a}{\xi^b}=\commut{\pi_a}{\pi_b}=0,\vs{6pt}\\
\commut{Z_{\alpha}}{Z_{\beta}}=i\hbar J^{\alpha\beta},
\end{array}
 \eqno{(2.4)}
$$ 
where $J^{\alpha\beta}$ is the $2N\times2N$ symplectic matrix.\\
\ts{12pt}We next define the symplectic \hope s $\Zpm_{\alpha}(\alpha=1,\dots,2M)$ as follows:\footnote{For any operators $A,B$, $\symmp{A}{B}=\dps{\f12}(AB+BA)$.}
$$
\Zm_{\alpha}=\f1{i\hbar}\commu{Z_{\alpha}}{\q},\hs{36pt}\Zp_{\alpha}=\symmp{Z_{\alpha}}{\q}.
\eqno{(2.5)}
$$
From (2.4), the \hope s $\Zpm$ obey the {\it hyper}-commutator algebra
$$
\begin{array}{l}
[\Zpm_{\alpha},\Zpm_{\beta}]=0,\vs{6pt}\\
\commu{\Zpm_{\alpha}}{\Zmp_{\beta}}=\commu{\Zmp_{\alpha}}{\Zpm_{\beta}}=J^{\alpha\beta}.
\end{array}
\eqno{(2.6)}
$$
Then, the projection operator $\Phat$ is given by
$$
\Phat=\exp\left[(-1)^s\Zp_{\alpha}\f{\partial}{\partial\varphi_{\alpha}}\right]\exp[J^{\alpha\beta}\varphi_{\alpha}\Zm_{\beta}]|_{\phi=0},
\eqno{(2.7)}
$$  
which stisfies the projection conditions 
$$
\Phat T_{\alpha}(\mcC)=T_{\alpha}(\Phat\mcC)=0\hs{48pt}(\alpha=1,\cdots,2M).
\eqno{(2.8)}
$$
\ts{12pt}The {\it hyper}-operator $\Qhat_{\eta\zeta}$ in the constraint star-products is defined by
$$
\Qhat_{\eta\zeta}=J^{\alpha\beta}\Zm_{\alpha}(\eta)\Zm_{\beta}(\zeta)=\hxi^{a}(\eta)\hpi_a(\zeta)-\hpi_a(\eta)\hxi^{a}(\zeta)
\eqno{(2.9)}
$$
with the nonlocal representations for the operations of {\it hyper}-operators, which satisfies
$$
\Qhat^t_{\eta\zeta}=\Qhat_{\zeta\eta}=-\Qhat_{\eta\zeta}.
\eqno{(2.10)}
$$
Then, we define two-kinds of star-product as follows:
$$
X\star Y=\left.\exp(\f{\hbar}{2i}\Omega_{\eta\zeta})X(\eta)Y(\zeta)\right|_{\eta=\zeta}
\eqno{(2.11)}
$$  
and
$$
X\pscrp\star   Y=\left.\left(\Phat(\eta)\Phat(\zeta)\exp(\f{\hbar}{2i}\Qhat^t_{\eta\zeta})X(\eta)Y(\zeta)\right)\right|_{\eta=\zeta}.
\eqno{(2.12)}
$$
\ts{12pt}Using the $\star$ and $\pscrp\star$-products, the commutator-formulas and the symmetrized product-ones under the operation of $\Phat$ are expressed as follows:
$$
\begin{array}{lcl}
\commu{\Phat X}{\Phat Y}&=&\Phat\commu{X}{Y}_{\star}=\Phat(X\star Y-Y\star X),\vs{12pt}\\
\symmp{\Phat X}{\Phat Y}&=&\Phat\symmp{X}{Y}_{\star}=\dps{\f12}\Phat(X\star Y+Y\star  X),
\end{array}
\eqno{(2.13a)}
$$  
and
$$
\begin{array}{lcl}
\Phat\commu{X}{Y}&=&\commu{X}{Y}_{\pscrp\star}=(X\pscrp\star Y-Y\pscrp\star X),\vs{12pt}\\
\Phat\symmp{X}{Y}&=&\symmp{X}{Y}_{\pscrp\star}=\dps{\f12}(X\pscrp\star Y+Y\pscrp\star X).
\end{array}
\eqno{(2.13b)}
$$

\section{Initial Hamiltonian System}

\ts{12pt}We shall propose the first-order singular model Lagrangian describing the noncommutative quantum system constrained to the curved space specified by $G(x)=0$ in $R^N$. Following the canonical quantization formulation for constraint systems\cite{A2,A9,A10}, we shall construct the unconstraint quantum system $\mcS$.

\subsection{Construction of Initial Hamitonian System $\mcS$}

\ts{12pt}Let $\Theta$ be the totally antisymmetric tensor defined by
$$
\Theta^{ij}=\theta\varepsilon^{ij},
\eqno{(3.1a)}
$$
where $\theta$ is the constant parameter with respect to the noncommutativity of coordinates, and $\varepsilon$, the completely antisymmetric tensor defined as
$$
\varepsilon^{ij}=1\hs{12pt}(i>j),\hs{12pt}\varepsilon^{ji}=-\varepsilon^{ij}\hs{24pt}(i,j=1,\cdots,N).
\eqno{(3.1b)}
$$
\ts{12pt}Consider, then, the dynamical system described by the first-order singular Lagrangian $L$
$$
\begin{array}{rcl}
L&=&L(x,\dot{x},v,\dot{v},\lambda,\dot{\lambda},u,\dot{u})\vs{12pt}\\

&=&\dot{x}^iv_i-\lambda\dot{G}(x)-\f12\dot{u}_i\Theta^{ij}u_j-h_0(x,v),
\end{array}
\eqno{(3.2)}
$$
where $\dot{G}(x)=\dot{x}^iG_i(x)$\footnote{$G_i(x)=\partial^x_{i}$ with $\partial^x_{i}= \partial/\partial x^i$.} and $h_0(x,v)$ corresponds to the Hamiltonian of the system,
$$
h_0(x,v)=\f12v_iv_i+V(x).
\eqno{(3.3)}
$$
\ts{12pt}Then, the initial unconstraint quantum system $\mcS=(\mcC,\mcA(\mcC),H(\mcC),\mcK)$ is obtained as follows: \vs{12pt}\\
{\bf \boldmath i) Initial canonically conjugate set $\mcC$}
$$ 
\ts{-72pt}\mcC=\{(x^i,p^x_i),(v_i,p_v^i),(\lambda,p_{\lambda}),(u_i,p_u^i)|i=1,\cdots,N\},
\eqno{(3.4)}
$$
which obeys the commutator algebra $\mcA(\mcC)$:
$$
\begin{array}{l}
\commut{x^i}{p^x_j}=i\hbar\delta^i_j,\hs{12pt}\commut{v_i}{p_v^j}=i\hbar\delta_i^j,\hs{12pt}\commut{u_i}{p_u^j}=i\hbar\delta_i^j,\vs{12pt}\\
\commut{\lambda}{p_{\lambda}}=i\hbar,\hs{12pt}
\mbox{(the others)}=0,
\end{array}
\eqno{(3.5)}
$$
{\bf \boldmath ii) Initial Hamiltonian $H$}
$$
H=\symmp{\mu^i_{(1)}}{\phi^{(1)}_i}+\symmp{\mu^i_{(2)}}{\phi^{(2)}_i}+\symmp{\mu_{(3)}}{\phi^{(3)}}+h_0(x,v),
\eqno{(3.6)}
$$
where $\phi^{(n)}$, $(n=1,\cdots, 3)$ are the constraint operators corresponding to the primary constraints together with $\phi^{(4)}_i$ $(i=1,\cdots,N)$ and $\mu^i_{(n)}$ $(n=1,\cdots,4)$ are the Lagrange multiplier operators.
\vs{6pt}\\
{\bf \boldmath iii) Consistent set of constraints and the Lagrange multiplier operators}
\vs{6pt}\\
\ts{12pt}Through the consistency conditions for the time evolusions of constraint operators, the consistent set of constraints, $\mcK$, is given as follows:
$$
\mcK=\{\phi^{\mbtn{(1)}}_i,\phi^{\mbtn{(2)}}_i,\phi^{\mbtn{(3)}},\phi^{\mbtn{(4)}}_i,\psi^{\mbtn{(1)}}\},
\eqno{(3.7)}
$$
where  
$$
\begin{array}{l}
\phi^{\mbtn{(1)}}_i=v_i-p^x_i-\lambda G_i(x),\vs{12pt}\\
\phi^{\mbtn{(2)}}_i=p_v^i,\vs{12pt}\\
\phi^{\mbtn{(3)}}=p_{\lambda},\vs{12pt}\\
\phi^{\mbtn{(4)}}_i=p_u^i+\dps{\f12}\Theta^{ij}u_j,
\end{array}
\eqno{(3.8a)}
$$
and
$$
\ts{-48pt}\psi^{\mbtn{(1)}}=G_i(x)v_i
\eqno{(3.8b)}
$$
which is the constraint operator corresponding to the secondary constraint.\\
\ts{12pt}The Lagrange multiplier operators, $\mu^i_{(1)}$, $\mu^i_{(2)}$, $\mu_{(3)}$ and $\mu^i_{(4)}$ are given by
$$
\begin{array}{l}
\mu_{\mbtn{(1)}}^i=-v_i,\vs{12pt}\\
\mu_{\mbtn{(2)}}^i=-P_{ij}(x)V_j(x)-G_i(x)\mcG^{-1}(x)G_{kl}(x)v_kv_l,\vs{12pt}\\
\mu_{\mbtn{(3)}}=\mcG^{-1}(x)(V_k(x)G_k(x)-G_{kl}(x)v_kv_l),\vs{12pt}\\
\mu_{\mbtn{(4)}}^i=0,
\end{array}
\eqno{(3.9)}
$$
where
$$
\begin{array}{l}
\mcG(x)=G_i(x)G_i(x),\vs{12pt}\\
P_{ij}(x)=\delta_{ij}-\mcG^{-1}(x)G_i(x)G_j(x),
\end{array}
\eqno{(3.10)}
$$
which satisfies
$$
P_{ij}(x)P_{jk}(x)=P_{ik}(x),\hs{36pt}P_{ij}(x)G_j(x)=G_i(x)P_{ij}(x)=0.
\eqno{(3.11)}
$$
\ts{12pt}The consistent set $\mcK$ obeys the commutator algebra $\mcA(\mcK)$:
$$
\begin{array}{l}
\commut{\phi^{\mbtn{(1)}}_i}{\phi^{\mbtn{(2)}}_j}=i\hbar\delta_{ij},\vs{12pt}\\
\commut{\phi^{\mbtn{(1)}}_i}{\phi^{\mbtn{(3)}}}=-i\hbar G_i(x),\vs{12pt}\\
\commut{\phi^{\mbtn{(1)}}_i}{\psi^{\mbtn{(1)}}}=i\hbar G_{ij}(x)v_j,\vs{12pt}\\
\commut{\phi^{\mbtn{(2)}}_i}{\psi^{\mbtn{(1)}}}=-i\hbar G_i(x),\vs{12pt}\\
\commut{\phi^{\mbtn{(4)}}_i}{\phi^{\mbtn{(4)}}_j}=i\hbar\Theta^{ij},\hs{48pt}\mbox{(the others)}=0.
\end{array}
\eqno{(3.12)}
$$
\ts{12pt}Thus, we have constructed the initial unconstraint quamtum system $\mcS$.

\section{Construction of Constraint Quantum System $\mcS^*$}

\ts{12pt}Starting with the initial system $\mcS$, we shall construct the constraint quantum system $\mcS^*$, which satisfies $\mcK=0$. \\
\ts{12pt}According to the structure of the commutator algebra (3.12), we classify $\mcK$ into the following two subsets  :
$$
\mcK=\mcK^{(\mbtn{A})}\oplus\mcK^{\mbtn{(B)}}\q \mbox{with}\q \mcK^{(\mbtn{A})}=\{\phi^{\mbtn{(1)}},\phi^{\mbtn{(2)}},\phi^{\mbtn{(3)}},\psi^{\mbtn{(1)}}\},\hs{12pt}\mcK^{\mbtn{(B)}}=\{\phi^{\mbtn{(4)}}\},
\eqno{(4.1)}
$$
where $\mcK^{(\mbtn{A})}$ and $\mcK^{\mbtn{(B)}}$ are commutable with each other.

\subsection{Projection of $\mcS$ to the system with $\mcK^{(\mbtn{A})}=0$}

\ts{12pt}The structure of $\mcA(\mcK^{(\mbtn{A})})$  is equivalent to that of $\mcA(\mcK)$ presented in the previous study\cite{A1}. As well as in our previous work\cite{A1}, therefore, the quantum system $\tmcS$ satisfying $\mcK^{(\mbtn{A})}=0$ is immediately constructed as follows:\\
\ts{12pt}Let $\Phat^{(\mbtn{1})}$  be the projection operator associated to the subset $\mcK^{(\mbtn{A})}$, that is, $\Phat^{(\mbtn{1})}\mcK^{(\mbtn{A})}=0.$ Then, $\tmcS$ is defined as
$$
\tmcS=\Phat^{(\mbtn{1})}\mcS=(\Phat^{(\mbtn{1})}\mcC,\Phat^{(\mbtn{1})}H,\Phat^{(\mbtn{1})}\mcK)=(\tmcC,\tH,\tmcK).
\eqno{(4.2)}
$$
{\bf \boldmath i) Projected CCS $\tmcC$}
\vs{6pt}\\
\ts{12pt}From the pojection condition $\Phat^{(\mbtn{1})}\mcK^{(\mbtn{A})}=0$, the projected CCS becomes
$$
\tmcC=\Phat^{(\mbtn{1})}\mcC=\{(x^i,p^x_i),(u_i,p_u^i)|i=1,\cdots,N\}
\eqno{(4.3a)}
$$
with
$$
\begin{array}{ll}
v_i=\symmp{P_{ij}(x)}{p^x_j},&p_v^i=0,\vs{6pt}\\
\lambda=-\symmp{\mcG^{-1}(x)G_i(x)}{p^x_i},&p_{\lambda}=0,
\end{array}
\eqno{(4.3b)}
$$
which obeys the commutator algebra $\mcA(\tmcC)$
$$
\begin{array}{lcl}
\mcA(\tmcC)&:&\commut{x^i}{p^x_j}=i\hbar\delta^i_j,\hs{12pt}\commut{u_i}{p_u^j}=i\hbar\delta^j_i,\vs{12pt}\\
& &\mbox{(the others)}=0.
\end{array}
\eqno{(4.4)}
$$
{\bf \boldmath ii) Projected constraints $\tmcK$ }
\vs{6pt}\\
\ts{12pt}Under the operation of $\Phat^{(\mbtn{1})}$, the constraint set $\mcK$ is transferred to $\tmcK$ as follows:
$$
\tmcK=\Phat^{(\mbtn{1})}\mcK=\Phat^{(\mbtn{1})}\mcK^{\mbtn{(B)}}=\mcK^{\mbtn{(B)}}=\{\phi^{\mbtn{(4)}}_i|i=1,\cdots,N\},
\eqno{(4.5a)}
$$
where
$$
\phi^{\mbtn{(4)}}_i=p_u^i+\f12\Theta^{ij}u_j.
\eqno{(4.5b)}
$$
{\bf  \boldmath iii) Projected Hamiltonian $\tH$}
\vs{6pt}\\
\ts{12pt}There arise the quantum correction terms in the projected Hamiltonian $\tH$ through the projections of $H$ and the re-ordering of operators in $\tH$\cite{A1}. Using the formulas of the POM, (2.13),  we obtain $\tH$ in the following form:
$$
\tH=\Phat^{(\mbtn{1})}H=\f12p^x_iP^{ij}(x)p^x_j+V(x)+U_1(x)+U_2(x).
\eqno{(4.6)}
$$
Here, $U_1(x)$ is the quantum correction caused by the projections of  $\dps{\f12}v_iv_i$ and $\symmp{\mu_{\mbtn{(3)}}}{\phi^{\mbtn{(3)}}}$ in $H$, which is given by
$$
U_1(x)=\f{3\hbar^2}8\mcG^{-2}(x)G_{ik}(x)G_k(x)G_{il}(x)G_l(x)=\f{3\hbar^2}{32}\mcG^{-2}(x)\mcG_{;i}(x)\mcG_{;i}(x)
\eqno{(4.7a)}
$$
with $\mcG_{;i}(x)=\partial^x_i\mcG(x)$, and $U_2(x)$ is caused by the re-ordering of operators in $\tH$,
$$
U_2(x)=\f{\hbar^2}4(n_{i;ij}(x)n_j(x)+n_{i;j}(x)n_{j;i}(x))+\f{\hbar^2}8(n_{i;i}(x)n_{j;j}(x)+n_{i;k}n_k(x)(x)n_{i;l}(x)n_l(x)),
\eqno{(4.7b)}
$$
where
$$
n_i(x)=\mcG^{-1/2}(x)G_i(x), \hs{24pt}n_{i;k,\cdots,l}(x)=\partial^x_k\cdots\partial^x_ln_i(x).
\eqno{(4.8)}
$$
{\bf  \boldmath iv) Projected Quantum System $\tmcS$}
\vs{6pt}\\
Thus, the quantum system $\tmcS$ is given by
$$
\tmcS=(\tmcC,\tH,\mcK^{(1)}),
\eqno{(4.9)}
$$
and it is easily shown that $\dot{G}(x)=0$  holds in $\tmcS$ as follows:
$$
\dot{G}(x)=\f1{i\hbar}\commut{G(x)}{\tH}=\symmp{p^x_i}{P_{ij}(x)G_j(x)}=0.
\eqno{(4.10)}
$$
In the projected system $\tmcS$, then, $G(x)$ is the constant of motion and the constraint $G(x)\approx 0$ is conserved through the time evolusion of the system\cite{A1,A11,A12}. 
 
 \subsection{Projection of  Projected Quantum System $\tmcS$}
 
 \ts{12pt}In order to eliminate the remaining constraint-subset $\mcK^{(\mbtn{1})}$, we shall impose the additional constraints upon  $\tmcS$, which produce the noncommutativity among the coordinates $x$ in the constraint quantum system $\mcS^*$.
 
 \subsubsection{ Additional Constraints}
 
\ts{12pt}Let $\psi^{\mbtn{(2)}}_i$ ($i=1,\cdots,N)$) be the additional constraint operators defined by
 $$
 \psi^{\mbtn{(2)}}_i=u_i-p^x_i\hs{36pt}(i=1,\cdots,N),
 \eqno{(4.11)}
 $$
 which correspond to the constraints 
 $$
  \psi^{\mbtn{(2)}}_i\approx 0.
  \eqno{(4.12)}
 $$
We shall denote the constraint-set consisting of $\phi^{\mbtn{(4)}}$ and $\psi^{\mbtn{(2)}}$ by $\mcK^{(\mbtn{2})}$:
$$
\mcK^{(\mbtn{2})}=\{\phi^{\mbtn{(4)}}_i,\psi^{\mbtn{(2)}}_i|i=1,\cdots,N\}.
\eqno{(4.13)}
$$
From the commutator algebra $\tmcC$, the constraint-set $\mcK^{(\mbtn{2})}$ obeys the commutator algebra $\mcA(\mcK^{(\mbtn{2})})$:
$$
\begin{array}{l}
\commut{\phi^{\mbtn{(4)}}_i}{\phi^{\mbtn{(4)}}_j}=i\hbar\Theta^{ij},\vs{12pt}\\
\commut{\psi^{\mbtn{(2)}}_i}{\phi^{\mbtn{(4)}}_j}=i\hbar\delta_{ij},\vs{12pt}\\
\commut{\psi^{\mbtn{(2)}}_i}{\psi^{\mbtn{(2)}}_j}=0,
\end{array}
\eqno{(4.14)}
$$
which shows that $\mcK^{(\mbtn{2})}$ is in the second-class. Therefore, the constraint quantum system $\mcS^*$ will be constructed through the star-product quantization formalism of POM\cite{A2,A4,A7}.
\vs{6pt}\\
{\bf  \boldmath i) ACCS}
\vs{6pt}\\
The ACCS associated with $\mcK^{(\mbtn{2})}$ is defined by 
$$
\{Z_{\alpha}|Z_{\alpha}\in\tmcC,\alpha=1,\cdots,2N\},
\eqno{(4.15)}
$$ 
where
$$
Z _{\alpha}=\left\{\begin{array}{l}\xi_i\hs{24pt}(\alpha=i)\vs{6pt}\\
\pi_i\hs{24pt}(\alpha=i+N) \hs{36pt}(\alpha=1,\cdots,2N\q;\q i=1,\cdots,N)
\end{array}\right.
\eqno{(4.16)}
$$
and
$$
\begin{array}{l}
\xi_i=\psi^{\mbtn{(2)}}_i=u_i-p^x_i,\vs{12pt}\\
\pi_i=\phi^{\mbtn{(4)}}_i-\dps{\f12}\Theta^{ij}\psi^{\mbtn{(2)}}_j=p_u^i+\dps{\f12}\Theta^{ij}p^x_j.
\end{array}
\eqno{(4.17)}
$$
Then, the projection operator $\Phat^{(\mbtn{2})}$ is represented in terms of $Z$, that is, $\Phat^{(\mbtn{2})}=\Phat^{(\mbtn{2})}(Z)$, which satisfies the projection conditions
$$
\Phat^{(\mbtn{2})}\psi^{\mbtn{(2)}}_i=0,\hs{12pt}\Phat^{(\mbtn{2})}\phi^{\mbtn{(4)}}_i=0\hs{24pt}(i=1,\cdots,N).
\eqno{(4.18)}
$$
\vs{6pt}\\
{\bf  \boldmath ii) The {\it hyper}-operator $\Qhat^{(\mbtn{2})}_{\eta\zeta}$}
\vs{6pt}\\
The {\it hyper}-operator $\Qhat^{(\mbtn{2})}_{\eta\zeta}$ in the star-products is obtained as follows:
$$
\begin{array}{lcl}
\Qhat^{(\mbtn{2})}_{\eta\zeta}&=&J^{\alpha\beta}\Zm_{\alpha}(\eta)\Zm_{\beta}(\zeta)=\hxi_i(\eta)\hpi_i(\zeta)-\hpi_i(\eta)\hxi_i(\zeta)\vs{12pt}\\
&=&-\hpxm{i}(\eta)\Theta^{ij}\hpxm{j}(\zeta)-\hpum{i}(\eta)\hum{i}(\zeta)+\hum{i}\hpum{i}(\zeta)\vs{12pt}\\
& &-\hpxm{i}(\eta)\hpum{i}(\zeta)+\hpum{i}(\eta)\hpxm{i}(\zeta)\vs{12pt}\\
& &+\dps{\f12}\Theta^{ij}(\hpxm{i}(\eta)\hum{i}(\zeta)+\hum{i}(\eta)\hpxm{i}(\zeta)).
\end{array}
\eqno{(4.19)}
$$

\subsubsection{Projection of $\tmcS$}

\ts{12pt}Let the projection of $\tmcS$ by $\Phat^{(\mbtn{2})}$ be $\mcS^{(\mbtn{2})}$. Then, the projected system  $\mcS^{(\mbtn{2})}$ is obtained in the following way:
$$
\mcS^{(\mbtn{2})}=\Phat^{(\mbtn{2})}\tmcS=(\Phat^{(\mbtn{2})}\tmcC,\Phat^{(\mbtn{2})}\tH,\Phat^{(\mbtn{2})}\mcK^{(\mbtn{2})})=(\mcC^{(\mbtn{2})},H^{(\mbtn{2})}).
\eqno{(4.20)}
$$ 
{\bf  \boldmath i) Projected CCS $\mcC^{(\mbtn{2})}$}
\vs{6pt}\\
From the projection conditions (4.18), the projection of $\tmcC$ becomes
$$
\mcC^{(\mbtn{2})}=\Phat^{(\mbtn{2})}\tmcC=\{x^i,p^x_i|i=1,\cdots,N\}
\eqno{(4.21)}
$$
with
$$
\begin{array}{l}
u_i=p^x_i,\hs{12pt}p_u^i=-\dps{\f12}\Theta^{ij}p^x_j,\vs{12pt}\\
v_i=\symmp{P_{ij}(x)^*}{p^x_j},\hs{12pt}\lambda=-\symmp{(\mcG(x)G_i(x))^*}{p^x_i},\vs{12pt}\\
p_v^i=0,\hs{12pt}p_{\lambda}=0,
\end{array}
\eqno{(4.22)}
$$
where, for any operator $A(x)\in\tmcS$, $A(x)^*=\Phat^{(\mbtn{2})}A(x)$. 
\vs{6pt}\\
{\bf  \boldmath ii) Commutator algebra of $\mcC^{(\mbtn{2})}$}
\vs{6pt}\\
\ts{12pt}From Eqs.(4.19) and (2.13), the commutator algebra $\mcA(\mcC^{(\mbtn{2})})$ is obtained as follows:
$$
\begin{array}{l}
\commut{x^i}{x^j}=i\hbar\Theta^{ij},\vs{12pt}\\
\commut{x^i}{p^x_j}=i\hbar\delta^i_j,\vs{12pt}\\
\commut{p^x_i}{p^x_j}=0.
\end{array}
\eqno{(4.23)}
$$
\vs{6pt}\\
{\bf  \boldmath iii) Projection of the Hamiltonian $\tH$}
\vs{6pt}\\
\ts{12pt}The projected Hamiltonian $H^{(\mbtn{2})}$ is given in the following form:
$$
H^{(\mbtn{2})}=\Phat^{(\mbtn{2})}\tH=\f12p^x_iP_{ij}(x)^*p^x_j+V(x)^*+U_1(x)^*+U_2(x)^*.
\eqno{(4.24)}
$$
As shown in Appendix A, then, the projected terms like $O(x)^*$ in (4.24) are  represented in the following form:
$$
O(x)^*=\Phat^*O(x)=\sum^{\infty}_{n=0}\hbar^{2n}O_n(x),
\eqno{(4.25)}
$$
 which contains the qunatum corrections in the form of the power series of $\hbar^{2n}$ $(n\geq 1)$\cite{A7}, where $x\in\mcC^{(\mbtn{2})}$ in $O_n(x)$ and $O_0(x)=O(x)$.
\vs{6pt}\\
{\bf  \boldmath iv) Constraint Quantum System $\mcS^*$}
\vs{6pt}\\
Through the successive projections of the initial system ,
$$
\Phat^{(\mbtn{1})}:\mcK^{(\mbtn{A})}=0 \rightarrow \Phat^{(\mbtn{2})}:\mcK^{(\mbtn{2})}=0,
\eqno{(4.26)}
$$
thus, we have constructed the resultant constraint quantum system 
$$
\mcS^*=(\mcC^*,H^*)=\mcS^{(\mbtn{2})}=(\mcC^{(\mbtn{2})}, H^{(\mbtn{2})}).
\eqno{(4.27)}
$$
As shown in Eqs.(4.24) and (4.25), the final Hamiltonian $H^*$ contains the quantum corretions in the form of the power series with $\hbar^{2n}$ in addition to those in $\tH$.

\subsection{Constraint Quantum System $\mcS^*_{\mbtn{QP}}$}

\ts{12pt}Following our previous works\cite{A2,A4}, we shall prepare the other additional constraints. Let $\Xi_{ij}$ be $\eta\varepsilon^{ij}$ with the noncommutativity constant parameter $\eta$, and let $G_{ij}$ be $
G_{ij}=(\Theta\Xi)_{ij}=(\Xi\Theta)_{ij}$ $(i,j=1,\cdots,N)$.

\subsubsection{Additional Constraints}

\ts{12pt}We consider the addtional constraint-operators\footnote{$\bM=I-\dps{\f14}G$}
$$
\psi^{\mbtn{(3)}}_i=\bM_{ij}u_j-p^x_i-\f12\Xi_{ij}x^j\hs{36pt}(i=1,\cdots,N),
\eqno{(4.28a)}
$$
which form the second-class constraint-set, $\mcK^{(\mbtn{2})}$, together with $\phi^{\mbtn{(4)}}_i$:
$$
\mcK^{(\mbtn{2})}=\{\phi^{\mbtn{(4)}}_i,\psi^{\mbtn{(3)}}_i|i=1,\cdots,N\}.
\eqno{(4.28b)}
$$
Then, the commutator algebra $\mcA(\mcK^{(\mbtn{2})})$ is given by
$$
\mcA(\mcK^{(\mbtn{2})})\e :\e \commut{\phi^{\mbtn{(4)}}_i}{\phi^{\mbtn{(4)}}_j}=i\hbar\Theta^{ij},\hs{12pt}\commut{\phi^{\mbtn{(4)}}_i}{\psi^{\mbtn{(3)}}_j}=i\hbar\bM_{ij},\hs{12pt}\commut{\psi^{\mbtn{(3)}}_i}{\psi^{\mbtn{(3)}}_j}=i\hbar\Xi_{ij}.
\eqno{(4.29)}
$$
\vs{6pt}\\
{\bf  \boldmath i) ACCS}
\vs{6pt}\\
\ts{12pt}The ACCS associated with $\mcK^{(\mbtn{2})}$ is defined as follows:
$$
Z_{\alpha}=\left\{\begin{array}{lcll}\xi_i&=&M^{-1}_{ij}(\tpsi^{\mbtn{(3)}}_j+\dps{\f12}\Xi_{jk}\tphi^{\mbtn{(4)}}_k)\vs{6pt}\\

&=&M^{-1}_{ij}(u_j+\dps{\f12}\Xi_{jk}p_u^k-p^x_j-\dps{\f12}\Xi_{jk}x^k)&\q (\alpha=i)\vs{12pt}\\

\pi_i&=&M^{-1}_{ij}(\tphi^{\mbtn{(4)}}_j-\dps{\f12}\Theta_{jk}\tpsi^{\mbtn{(3)}}_k)\vs{6pt}\\

&=&M^{-1}_{ij}(p_u^j+\dps{\f18}(G\Theta)_{jk}u_k+\dps{\f12}\Theta^{jk}p^x_k+\dps{\f14}G_{jk}x^k)&\q (\alpha=i+N),\end{array}\right.
\eqno{(4.30)}
$$
\vs{6pt}\\
{\bf  \boldmath ii) The {\it hyper}-operator $\Qhat^{(\mbtn{2})}_{\eta\zeta}$}
\vs{6pt}\\
Then, the {\it hyper}-operator $\Qhat^{(\mbtn{2})}_{\eta\zeta}$ is obtained as follows: 
$$
\Qhat^{(\mbtn{2})}_{\eta\zeta}=J^{\alpha\beta}\Zm_{\alpha}(\eta)\Zm_{\beta}(\zeta)=\hxi^{(-)}_i(\eta)\hpi^{(-)}_i(\zeta)-\hpi^{(-)}_i(\eta)\hxi^{(-)}_i(\zeta)
\eqno{(4.31a)}
$$
with
$$
\begin{array}{lclclcl}

\hxi^{(-)}_kx^i&=&M^{-1}_{ki},& &\hpi^{(-)}_kx^i&=&-\dps{\f12}(M^{-1}\Theta)_{ki},\vs{6pt}\\

\hxi^{(-)}_kp^x_i&=&-\dps{\f12}(M^{-1}\Xi)_{ki},& &\hpi^{(-)}_kp^x_i&=&\dps{\f14}(M^{-1}G)_{ki},\vs{6pt}\\

\hxi^{(-)}_ku_i&=&-\dps{\f12}(M^{-1}\Xi)_{ki},& &\hpi^{(-)}_ku_i&=&-M^{-1}_{ki}.

\end{array}
\eqno{(4.31b)}
$$

\subsubsection{Projected Quantum System $\mcS^{(\mbtn{2})}$}

\ts{12pt}Through the constraint star-product quantization formalism of POM\cite{A2,A4,A7}, we obtain the projected quantum system $\mcS^{(\mbtn{2})}$ in the following way:
$$
\mcS^{(\mbtn{2})}=\Phat^{(\mbtn{2})}\tmcS=(\Phat^{(\mbtn{2})}\tmcC,\Phat^{(\mbtn{2})}\tH,\Phat^{(\mbtn{2})}\mcK^{(\mbtn{2})})=(\mcC^{(\mbtn{2})},H^{(\mbtn{2})}),
\eqno{(4.32)}
$$
where $\Phat^{(\mbtn{2})}$ is the projection operator associated to the ACCS (4.30).
\vs{6pt}\\
{\bf  \boldmath i) Projected CCS $\mcC^{(\mbtn{2})}$}
\vs{6pt}\\
\ts{12pt}The projected CCS $\mcC^{(\mbtn{2})}$ is obtained as follows: 
$$
\mcC^{(\mbtn{2})}=\Phat^{(\mbtn{2})}\tmcC=\{(x^i,p^x_i),(u_i,p_u^i)|i=1,\cdots,N\}
\eqno{(4.33a)}
$$
with
$$
\begin{array}{lcl}

p_u^i+\dps{\f12}\Theta^{ij}p^x_j=0,& &\bM_{ij}u_j-p^x_i-\dps{\f12}\Xi_{ij}x^i=0,\vs{12pt}\\

v_i=\symmp{P_{ij}(x)^*}{p^x_j},&  &p_v^i=0,\vs{12pt}\\

\lambda=-\symmp{(\mcG(x)G_i(x))^*}{p^x_i},& &p_{\lambda}=0.

\end{array}
\eqno{(4.33b)}
$$
\vs{6pt}\\
{\bf  \boldmath ii) Commutator algebra of $\mcC^{(\mbtn{2})}$}
\vs{6pt}\\
\ts{12pt}Using Eqs.(4.31) in the projection formulas (2.13), then, the commutator algebra of $\mcA(\mcC^{(\mbtn{2})})$ is given by 
$$
\begin{array}{lcl}

\commut{x^i}{x^j}=i\hbar(M^{-1}\Theta M^{-1})_{ij},&  &\commut{u_i}{u_j}=i\hbar(M^{-1}\Xi M^{-1})_{ij},\vs{12pt}\\

\commut{x^i}{p^x_j}=i\hbar(M^{-1}(I+\dps{\f1{16}}G^2)M^{-1})_{ij},&  &\commut{u_i}{p_u^j}=i\hbar\dps{\f12}(M^{-1}GM^{-1})_{ij},\vs{12pt}\\

\commut{p^x_i}{p^x_j}=-i\hbar\dps{\f14}(M^{-1}G\Xi M^{-1})_{ij},&  &\commut{p_u^i}{p_u^j}=-i\hbar\dps{\f14}(M^{-1}G\Theta M^{-1})_{ij},\vs{12pt}\\

\commut{x^i}{u_j}=i\hbar(M^{-1}\bM M^{-1})_{ij},& &\commut{u_i}{p^x_j}=i\hbar\dps{\f12}(M^{-1}\Xi\bM M^{-1})_{ij},\vs{12pt}\\

\commut{x^i}{p_u^j}=i\hbar\dps{\f12}(M^{-1}\Theta\bM M^{-1})_{ij},& &\commut{p^x_i}{p_u^j}=i\hbar\dps{\f14}(M^{-1}G\bM M^{-1})_{ij}.

\end{array}
\eqno{(4.34)}
$$ 
which shows manifestly that $\mcS^{(\mbtn{2})}$ satisfies the noncommutativities both the coordinates and the momentum operators.
\vs{6pt}\\
{\bf  \boldmath iii) Projected Hamiltonian $H^{(\mbtn{2})}$}
\vs{6pt}\\
\ts{12pt}From Eqs.(4.33) and (4.34), then, the projected Hamiltonian $H^{(\mbtn{2})}$  is expressed in the following two forms:
$$
H^{(\mbtn{2})}=\left\{
\begin{array}{l}
H^{(\mbtn{2})}(x,p)=\dps{\f12}p^x_iP_{ij}(x)^*p^x_j+U_c(x)^*+V(x)^*+U_1(x)^*+U_2(x)^*,\vs{12pt}\\

H^{(\mbtn{2})}(x,u)=\dps{\f12}u_iP_{ij}(x)^*u_j+A_{\mbtn{K}}(x,u)+U_c(x)^*+V(x)^*+U_1(x)^*+U_2(x)^*,

\end{array}\right.
\eqno{(4.35a)}
$$
where $A_{\mbtn{K}}(x,u)$ is the additional term caused by representing $p^x$ in terms of $u$, which is expressed as
$$
A_{\mbtn{K}}(x,u)=\f14\bM_{ik}\Xi_{jl}(u_iP_{kl}(x)^*x^j+x^jP_{kl}(x)^*u_i)-\f14x^i\Xi_{ik}P_{kl}(x)^*\Xi_{lj}x^j.
\eqno{(4.35b)}
$$
Here, $U_c(x)^*$ is the quantum correction term caused by the projection of $\dps{\f12}p^x_iP_{ij}(x)p^x_j$, which is given by 
$$
U_c(x)^*=\dps{\f{\hbar^2}8}(M^{-1}GM^{-1})_{ij}P_{ik;jk}(x)^*-\dps{\f{\hbar^2}{32}}(M^{-1}GM^{-1})_{ij}P_{ik;jl}(x)^*(M^{-1}GM^{-1})_{kl},
\eqno{(4.36)}
$$
where $P_{i\cdots j;k\cdots l}(x)^*=\Phat^{(\mbtn{2})}P_{i\cdots j;k\cdots l}(x)$ and, from the commutator algebra (4.34), the projected terms in $H^{(\mbtn{2})}$ are calculated through replacing $\Theta^{ij}$ by $M^{-1}\Theta^{ij}M^{-1}$. 

\subsubsection{Constraint Quantum System $\mcS^*_{\mbtn{QP}}$}

\ts{12pt}From the projection conditions (4.33b) and the commutator algebra (4.34) in $\mcS^{(\mbtn{2})}$, two kinds of constraint quantum system are set up in the construction of $\mcS^*_{\mbtn{QP}}$, which we shall denote by $\mcS^*_{\mbtn{I}}$ and $\mcS^*_{\mbtn{II}}$, respectively. 
\vs{6pt}\\
{\bf  \boldmath i) Constraint Quantum System $\mcS^*_{\mbtn{I}}$}
\vs{6pt}\\
\ts{12pt}The constraint quantum system $\mcS^*_{\mbtn{I}}$ is constructed with the CCS $\{(x^i,u_i)|i=1,\cdots,N\}$ in the following way: 
$$
\mcS^*_{\mbtn{I}}=(\mcC^*_{\mbtn{I}},H^*_{\mbtn{I}}).
\eqno{(4.37)}
$$
Here,
$$
\mcC^*_{\mbtn{I}}=\{(x^i,u_i)|i=1,\cdots,N\}
\eqno{(4.38a)}
$$
with
$$
\begin{array}{lcl}

p^x_i=\bM_{ij}u_j-\dps{\f12}\Xi_{ij}x^j,&  &p_u^i=-\dps{\f12}\Theta^{ij}u_j,\vs{12pt}\\

v_i=\symmp{P_{ij}(x)^*}{\bM_{jk}u_k-\dps{\f12}\Xi_{jk}x^k},&  &p_v^i=0,\vs{12pt}\\

\lambda=-\symmp{(\mcG(x)G_i(x))^*}{\bM_{jk}u_k-\dps{\f12}\Xi_{jk}x^k},& &p_{\lambda}=0.

\end{array}
\eqno{(4.38b)}
$$
Then, the commutator algebra $\mcA(\mcC^*_{\mbtn{I}})$ is defined as
$$
\begin{array}{ll}

\mcA(\mcC^*_{\mbtn{I}}):

&\commut{x^i}{x^j}=i\hbar(M^{-1}\Theta M^{-1})_{ij},\vs{12pt}\\

&\commut{x^i}{u_j}=i\hbar(M^{-1}\bM M^{-1})_{ij},\vs{12pt}\\

&\commut{u_i}{u_j}=i\hbar(M^{-1}\Xi M^{-1})_{ij},

\end{array}
\eqno{(4.39)}
$$
and the resultant Hamiltonian $H^*_{\mbtn{I}}$ is given by
$$
H^*_{\mbtn{I}}=H^{(2)}(x,u)=\dps{\f12}u_iP_{ij}(x)^*u_j+A_{\mbtn{K}}(x,u)+U_c(x)^*+V(x)^*+U_1(x)^*+U_2(x)^*.
\eqno{(4.40)}
$$
\vs{6pt}\\
{\bf  \boldmath ii) Constraint Quantum System $\mcS^*_{\mbtn{II}}$}
\vs{6pt}\\
\ts{12pt}As well as in $\mcS^*_{\mbtn{I}}$, the constraint quantum system  $\mcS^*_{\mbtn{II}}$ is constructed in the following way:
$$
\mcS^*_{\mbtn{II}}=(\mcC^*_{\mbtn{II}},H^*_{\mbtn{II}}),
\eqno{(4.41)}
$$
where
$$
\mcC^*_{\mbtn{II}}=\{(x^i,p^x_i|i=1,\cdots,N\}
\eqno{(4.42a)}
$$
with
$$
\begin{array}{lcl}

u_i=\bM^{-1}_{ij}(p^x_j+\dps{\f12}\Xi_{jk}x^k),&  &p_u^i=-\dps{\f12}\Theta^{ij}u_j=-\dps{\f12}(\Theta\bM^{-1})_{ij}(p^x_j+\dps{\f12}\Xi_{jk}x^k),\vs{12pt}\\

v_i=\symmp{P_{ij}(x)^*}{p^x_j},&  &p_v^i=0,\vs{12pt}\\

\lambda=-\symmp{(\mcG(x)G_i(x))^*}{p^x_j},& &p_{\lambda}=0,

\end{array}
\eqno{(4.42b)}
$$
of which the commutator algebra $\mcA(\mcC^*_{\mbtn{II}})$ is
$$
\begin{array}{ll}

\mcA(\mcC^*_{\mbtn{II}}):

&\commut{x^i}{x^j}=i\hbar(M^{-1}\Theta M^{-1})_{ij},\vs{12pt}\\

&\commut{x^i}{p^x_j}=i\hbar(M^{-1}(\dps{I+\f1{16}G^2})M^{-1})_{ij},\vs{12pt}\\

&\commut{p^x_i}{p^x_j}=-i\hbar\dps{\f14}(M^{-1}G\Xi M^{-1})_{ij},

\end{array}
\eqno{(4.43)}
$$
and $H^*_{\mbtn{II}}$ is the resultant Hamiltonian
$$
H^*_{\mbtn{II}}=H^{(2)}(x,p^x)=\dps{\f12}p^x_iP_{ij}(x)^*p^x_j+U_c(x)^*+V(x)^*+U_1(x)^*+U_2(x)^*.
\eqno{(4.44)}
$$

\section{Discussion and Concluding remarks}
\ts{12pt}In order to construct the noncommutative quantum system on the curved space {\it exactly}, we have proposed the Lagrangian $L$ with the {\it dynamical} constraint, which has been obtained by modifying the first-order singular Lagrangians in noncommutative quantum theories\cite{A2,A13} through adding the redundant variables $u$. \\
\ts{12pt}Starting with the Lagrangian $L$, we have at first composed the quantum system $\tmcS$ constrained to the curved space $M^{N-1}$. Imposing the additional constraints (4.11) upon $\tmcS$, we have next constructed the constraint quantum system $\mcS^*$ with the noncommutativity among $x$,
$$
\commut{x^i}{x^j}=i\hbar\Theta^{ij},
\eqno{(5.1)}
$$
on the curved space.  We have then shown that the resultant system $\mcS^*$ is expressed with $\mcC^*=\{(x,p^x)\}$ and the Hamiltonian $H^*$ contains the quantum correction terms in the form of the power-series of  $\hbar^{2n}$ $(n\geq 1)$, which are completely missed in the usual approach with the Dirac-bracket quantization\cite{A9,A10}.\\
\ts{12pt}Imposing the other additional constraints (4.28), further, we have constructed the two kinds of constraint quantum systems, $\mcS^*_{\mbtn{I}}$ and $\mcS^*_{\mbtn{II}}$ as the resultant constraint quantum system $\mcS^*_{\mbtn{QP}}$. Then, we have shown that the CCS's, $\mcC^*_{\mbtn{I}}$ and $\mcC^*_{\mbtn{I}}$, of these two systems are noncommutative about both of the coordinates $x$ and the momentum operators $p^x$, $u$, which obey the commutator algebras (4.39), (4.43). We have also shown that the resultant system $\mcS^*_{\mbtn{QP}}$ is expressed with the resultant Hamiltonian $H^*_{\mbtn{QP}}(=H^{(\mbtn{2})})$ (4.38b) contains the quantum corrections in the similar form to those in $H^*$.\\
\ts{12pt}We have thus constructed the noncommutative quantum systems on a curved space in the {\it exact} form. 
\vs{12pt}\\

\appendix

\ts{-24pt}{\bf \LARGE Appendix}

\section{Projection of the product of operators by $\Phat^*$}

By the use of the star-product quantization formulas of POM, Eqs.(2.13),  with the {\it hyper}-operator (4.19), the projection of the product of any two operators $X(x),Y(x)\in\tmcS$ by $\Phat^*$ is given as follows:
$$
\begin{array}{l}

\Phat^*(X(x)Y(x))=\Phat^*\symmp{X(x)}{Y(x)}=\dps{\f12}\commut{X(x)}{Y(x)}_+\vs{12pt}\\

=\dps{\f12\sum^{\infty}_{n=0}}\left(\dps{\f{\hbar}2}\right)^{2n}\dps{\f{(-1)^n}{(2n)!}}\Theta^{2n}_{[kl]}\commut{\Phat^*X_{;k_{2n}\cdots k_1}(x)}{\Phat^*Y_{;l_{2n}\cdots l_1}(x)}_+\vs{12pt}\\

+\dps{\f12\sum^{\infty}_{n=0}}\left(\dps{\f{\hbar}2}\right)^{2n+1}\dps{\f{(-1)^n}{(2n+1)!}}\Theta^{2n+1}_{[kl]}(1/i)\commut{\Phat^*X_{;k_{2n+1}\cdots k_1}(x)}{\Phat^*Y_{;l_{2n+1}\cdots l_1}(x)},

\end{array}
\eqno{(\mbox{A}1)}
$$ 
where
$$
\Theta^n_{[kl]}=\Theta^{k_nl_n}\cdots\Theta^{k_1l_1},
\eqno{(\mbox{A}2)}
$$
$$
\commut{X}{Y}_+=XY+YX,
\eqno{(\mbox{A}3)}
$$ 
$X_{;k_n\cdots k_1}(x)=\partial^x_{k_n}\cdots\partial^x_{k_1}X(x)$ and $Y(x)$, so also.\\ 
\ts{12pt}It is noticed that the expression (A1) is the Hermitian for the Hermite operators $X(x)$ and $Y(x)$.\\
\ts{12pt}From the formula (A1), the projection of $X^2(x)$ is given by
$$
\Phat^*X^2(x)=\Phat^*\symmp{X(x)}{X(x)}
=\dps{\sum^{\infty}_{n=0}}\left(\dps{\f{\hbar}{2i}}\right)^n\dps{\f1{n!}}\Theta^n_{[kl]}(\Phat^*X_{;k_n\cdots k_1}(x))(\Phat^*X_{;l_n\cdots l_1}(x)).
\eqno{(\mbox{A}4)}
$$
Then, the quantum corrections $U_1(x)^*$ and $U_2(x)^*$ in $H^*$ are given as follows:
$$
\begin{array}{lcl}

U_1(x)^*&=&\Phat^*U_1(x)=\dps{\f{3\hbar^2}{32}}\Phat^*(\mcG^{-1}(x)\mcG_{;i}(x))^2\vs{12pt}\\

&=&\dps{\f{3\hbar^2}{32}\sum^{\infty}_{n=0}\left(\f{\hbar}{2i}\right)^n\f1{n!}}\Theta^n_{[kl]}(\mcG^{-1}(x)\mcG_{;i}(x))^*_{;k_n\cdots k_1}(\mcG^{-1}(x)\mcG_{;i}(x))^*_{;l_n\cdots l_1}.

\end{array}
\eqno{(\mbox{A}5)}
$$ 
$$
\begin{array}{lcl}

U_2(x)^*&=&\Phat^*U_2(x)\vs{12pt}\\

&=&\dps{\f{\hbar^2}8}\dps{\sum^{\infty}_{n=0}\left(\f{\hbar}2\right)^{2n}\f{(-1)^n}{(2n)!}}\Theta^{2n}_{[kl]}\vs{6pt}\\

& &\times(\commut{n_{i;jk_{2n}\cdots k_1}(x)^*}{n_{j;il_{2n}\cdots l_1}(x)^*}_++\commut{n_{i;ijk_{2n}\cdots k_1}(x)^*}{n_{j;l_{2n}\cdots l_1}(x)^*}_+)\vs{12pt}\\

& &\ts{-12pt}+\dps{\f{\hbar^2}8}\dps{\sum^{\infty}_{n=0}\left(\f{\hbar}2\right)^{2n+1}\f{(-1)^n}{(2n+1)!}}\Theta^{2n+1}_{[kl]}\vs{6pt}\\

& &\times(1/i)(\commut{n_{i;jk_{2n+1}\cdots k_1}(x)^*}{n_{j;il_{2n+1}\cdots l_1}(x)^*}+\commut{n_{i;ijk_{2n+1}\cdots k_1}(x)^*}{n_{j;l_{2n+1}\cdots l_1}(x)^*})\vs{12pt}\\

& &\ts{-12pt}+\dps{\f{\hbar^2}8}\dps{\sum^{\infty}_{n=0}\left(\f{\hbar}{2i}\right)^n\f1{n!}}\Theta^n_{[kl]}\vs{6pt}\\

& &\times(n_{i;ik_n\cdots k_1}(x)^*n_{j;jl_n\cdots l_1}(x)^*+(n_{i;k}(x)n_k(x))^*_{;k_n\cdots k_1}(n_{i;l}(x)n_l(x))^*_{;l_n\cdots l_1}.

\end{array}
\eqno{(\mbox{A}6)}
$$
Using (A1) and (A2) repeatedly, then, $U_{\alpha}(x)^*$ ($\alpha=1,2$) are represented in the following form:
$$
U_{\alpha}(x)^*=\sum^{\infty}_{n=0}\hbar^{2n}U_{\alpha}^{(n)}(x)
\eqno{(\mbox{A}7)}
$$
with $x\in\mcS^*$, and, $P_{ij}(x)^*$, $V(x)^*$ in $H^*$, so also.

\end{document}